\documentclass{emulateapj}

\usepackage{times}

\newcommand{\etal}{{et al.}\/ }
\newcommand{\eg}{{\em e.g. }}
\newcommand{\ie}{{\em i.e. }}
\newcommand{\vs}{{vs. }}
\newcommand{\msun}{\hbox{M$_{\odot}$}}
\newcommand{\Lsun}{\hbox{L$_{\odot}$}}

\newcommand{\rxj}{RX~J0142.0+2131}

\newcommand{\cosmology}{$H_0 = 70$ km s$^{-1}$ Mpc$^{-1}$,
$\Omega_{\mathrm{m}} = 0.3$, $\Omega_{\Lambda} = 0.7$}

\newcommand{\afe}{$[\alpha/\mathrm{Fe}]$}
\newcommand{\zform}{$z_{\mathrm{form}}$}

\newcommand{\re}{r_{\mathrm{e}}}
\newcommand{\Ie}{\langle I \rangle_{\mathrm{e}}}

\shorttitle{\rxj: the Fundamental Plane} 
\shortauthors{Barr et al.}

\begin{document}

\title{The Fundamental Plane in \rxj: a galaxy cluster merger at
\lowercase{$z=0.28$}}

\author{
Jordi Barr\altaffilmark{1}, 
Inger J{\o}rgensen\altaffilmark{2},
Kristin Chiboucas\altaffilmark{2},
Roger Davies\altaffilmark{1},
Marcel Bergmann\altaffilmark{3}
}

\altaffiltext{1}{Oxford University, Denys Wilkinson Building, Keble
Road, Oxford, OX1 3RH, UK; jmb@astro.ox.ac.uk, rld@astro.ox.ac.uk}

\altaffiltext{2}{Gemini Observatory, 670 N A`ohoku Pl., Hilo, HI
96720; ijorgensen@gemini.edu, kchibouc@gemini.edu}

\altaffiltext{3}{Gemini Observatory, Casilla 603, La Serena, Chile;
mbergmann@gemini.edu}

\begin{abstract}

We present the Fundamental Plane (FP) in the $z = 0.28$ cluster of
galaxies \rxj. There is no evidence for a difference in the slope of
the FP when compared with the Coma cluster, although the internal
scatter is larger. On average, stellar populations in \rxj \ have
rest-frame $V$-band mass-to-light ratios ($M/L_V$) $0.29 \pm 0.03$ dex
lower than in Coma. This is significantly lower than expected for a
passively-evolving cluster formed at $z_{f} = 2$. Lenticular galaxies
have lower average $M/L_V$ and a distribution of $M/L_V$ with larger
scatter than ellipticals. Lower mass-to-light ratios are not due to
recent star formation: our previous spectroscopic observations of \rxj
\ E/S0 galaxies showed no evidence for significant star-formation
within the past $\sim 4$ Gyr. However, cluster members have enhanced
$\alpha$-element abundance ratios, which may act to decrease
$M/L_V$. The increased scatter in the \rxj \ FP reflects a large
scatter in $M/L_V$ implying that galaxies have undergone bursts of
star formation over a range of epochs. The seven easternmost cluster
galaxies, including the second brightest member, have $M/L_V$
consistent with passive evolution and $z_{f} = 2$. We speculate that
\rxj \ is a cluster--cluster merger where the galaxies to the east are
yet to fall into the main cluster body or have not experienced star
formation as a result of the merger.

\end{abstract}


\section{Introduction}

The Fundamental Plane (FP) of elliptical (E) and lenticular (S0)
galaxies is a log-linear relation between effective radius, surface
brightness, and velocity dispersion (\eg Djorgovski \& Davis 1987;
Dressler \etal 1987; J{\o}rgensen \etal 1996, hereafter
JFK1996)\nocite{djorgovski87,dressler87,jorgensen96}. Its ubiquity in
nearby clusters of galaxies has been noted by many investigators (\eg
Bender \etal 1992; JFK1996; Colless \etal
2001)\nocite{bender92,jorgensen96,colless01}.  Coefficients in the
observed FP differ from those predicted by the virial theorem, and
this `tilt' implies that mass-to-light ratio ($M/L$) is not a constant
for E/S0s but depends to some extent on galaxy properties (see
Cappellari \etal 2006\nocite{cappellari06}).  The FP must confront
studies showing that the growth of galaxies in clusters involves
complex processes like galaxy mergers, cluster mergers, and AGN
activity (\eg Fabian \etal 2000; Tran \etal
2005)\nocite{fabian00,tran05}.
 
If E/S0 galaxies form in a short burst and their stellar populations
subsequently evolve passively, one would expect a gradual decrease in
$M/L$ with cosmic epoch.  This induces an offset in the FP as a
function of redshift, with no change in slope. However, the tendency
of star formation to occur over a longer period in lower-mass galaxies
(\eg Juneau \etal 2005; Treu \etal 2005)\nocite{juneau05,treu05},
would be manifest as a steepening of the empirical FP with
redshift. Evidence for changes in the slope of the FP are seen in the
latest studies of the FP at $z>0.8$ (\eg Holden \etal 2005;
J{\o}rgensen \etal 2006)\nocite{holden05,jorgensen06}, but this has
not been observed at $0.2 < z < 0.8$ (\eg van Dokkum \& Franx 1996;
Kelson \etal 2000; Wuyts \etal
2004)\nocite{vandokkum96,kelson00b,wuyts04}, most likely because these
studies do not probe far enough down the mass function. The internal
scatter of the FP reported at $0.2 < z < 0.8$ is consistent with that
of the Coma cluster (JFK1996)\nocite{jorgensen96}. Since scatter in
the FP is a reflection of the variation in $M/L$, this is taken as
evidence that cluster E/S0s at a given mass are coeval.

In this letter we examine the FP in \rxj, a cluster of galaxies at
$z=0.28$ for differences in slope and internal scatter with respect to
that of Coma. We use \cosmology, and convert previous work to these
values where necessary. The look-back time to \rxj \ in this cosmology
is 3.2 Gyr.

\section{\rxj : Previous observations}
\label{sec:rxj}

We presented high--signal-to-noise GMOS-N spectroscopy of 43
spectroscopic targets in the field of \rxj \ in Barr \etal (2005;
hereafter BDJBC)\nocite{barr05}. Velocity dispersions and line indices
for 30 cluster members were derived. We determined that scaling
relations between metal indices and velocity dispersion, and the
strengths of the 4000\AA \ break, are inconsistent with a scenario in
which stellar populations form at $z_f \gtrsim 2$ and evolve passively
to look like cluster galaxies at low redshift (\ie without additional
star formation and/or merging). Stellar populations in \rxj \ have
$\alpha$-element abundance ratios (\afe) which are enhanced by, on
average, $0.14 \pm 0.03$ over Coma galaxies. Luminosity-weighted mean
ages for galaxies in \rxj \ are similar to those in Coma, \ie older
than would be expected at $z = 0.28$. The cluster velocity dispersion
is much larger than its X-ray luminosity or richness suggest. Despite
this, no sign of substructure was found by testing the distribution of
spectroscopically-confirmed cluster members in RA,dec,$z$ space.

\section{Data reduction and analysis}
\label{sec:obs}

The data reduction and analysis method for the GMOS-N spectroscopy was
presented in BDJBC\nocite{barr05}. In this letter we restrict the
surface photometry calculations to spectroscopically-confirmed cluster
members. We will present effective radii, mean surface brightnesses
and morphological classifications for a larger sample at a later date.

\rxj \ was observed with the Hubble Space Telescope Advanced Camera
for Surveys (ACS) on UT 2003 November 01 and 2004 July 03. Two
positions were imaged in the F775W band for 4420s each. Reductions are
performed with the PyRAF task MULTIDRIZZLE using the standard
procedure. ACS surface photometry is analysed using the
two-dimensional fitting program GALFIT \cite{peng02}. We fit galaxies'
luminosity profiles as as $r^{1/4}$ profiles.  Background or companion
galaxies may influence the derived parameters, so these are fit
simultaneously as S{\'e}rsic profiles \cite{sersic68}. We also
classify galaxies by Hubble type using a method analogous to that of
Smail \etal (1997)\nocite{smail97}. Magnitudes and surface
brightnesses are corrected to rest-frame $V$ using F775W magnitudes
and GMOS colors in a similar way as used in BDJBC\nocite{barr05}. A
complete description of our data, reduction techniques, analysis of
potential systematics as well as our method of deriving
surface-brightness parameters from the GALFIT fitting program will be
presented in a future paper (Chiboucas et al., in
preparation). Twenty-eight spectroscopically-confirmed cluster members
were covered by the two ACS fields. Of these 11 are E, 12 are S0 and
the remaining 5 are spirals, irregulars, mergers, or not classifiable.

We estimate the systematic and random errors in the values returned by
GALFIT using Monte Carlo simulations. Artificial galaxies are
generated with GALFIT and with the IRAF routine
ARTDATA/MKOBJECTS. These objects cover the full range of parameter
space in magnitude, $\re$, axis ratio,and position angle of the \rxj \
sample. The S{\'e}rsic index is varied randomly between 2.5 and 5 and
we adjust the diskiness/boxiness of each artificial galaxy. Noise is
then added to the artificial galaxies which are placed randomly in the
ACS images and recovered as $r^{1/4}$ profiles. 

The results of the simulations indicate that $\log \re$ derived for
the $r^{1/4}$ profile is systematically greater than the simulated
value by 0.06 dex. The quantity, $\log \re + \beta \log \Ie$, which
enters the FP and referred to hereafter as the Fundamental Plane
Parameter (FPP), is known to be much less sensitive to the imposition
of the $r^{1/4}$-law fit (\eg Lucey 1997)\nocite{lucey97}. The derived
FPP is systematically offset from its simulated value by $-0.02$, with
a standard deviation of $0.03$. The magnitude of the deviation is not
a function of the FPP. We do not make corrections to the surface
photometry but quantify the total uncertainty on the FPP as $0.05$.

\section{The comparison sample}
\label{sec:com}

Our comparison sample consists of 116 E/S0 galaxies in the Coma
cluster; velocity dispersions are from J{\o}rgensen
(1999)\nocite{jorgensen99}. We use Gunn $r$ surface photometry of E/S0
galaxies fit using an $r^{1/4}$ law from J{\o}rgensen \etal
(1995)\nocite{jorgensen95}, as well as new $B$ and $R_c$ photometry
obtained with the McDonald Observatory 0.8m telescope. We transform
from Gunn $r$ to the rest-frame $V$ using $r -R_c = 0.354$
\cite{jorgensen94}, $V = R_c + 0.337(B - R_c) + 0.089$ derived for
E/S0 galaxies using Bruzual \& Charlot (2003)\nocite{bruzual03}
models, a median Coma red-sequence color of $B - R_c = 1.485$, and
$V_{\mathrm{rest}} = V - Q_V - 10 \log (1 + z)$ where the
$k$-correction, $Q_V = 0.043$, is derived from Bruzual \& Charlot
(2003)\nocite{bruzual03} models. The surface brightness parameter in
the $V$-band FP is $\log \Ie = -0.4(\langle \mu \rangle_{e} - 26.43)$.

We recalculate the FP for the Coma cluster by minimising the absolute
residuals in $\log \re, \log \sigma, \log \Ie$ perpendicular to the
plane, following JFK1996\nocite{jorgensen96}. Three emission-line
galaxies are excluded from the fit. We find 
\[
\log \re = (1.29 \pm 0.07) \log \sigma - (0.83 \pm 0.03) \log \Ie - 0.28\label{equ:fpc}
\]
with an rms of 0.08 in $\log r_{e}$.

\section{The Fundamental Plane in \rxj}
\label{sec:fpr}

We fit the FP to E/S0 galaxies in \rxj \ in the same way as for the
Coma cluster. Emission-line galaxies identified in
BDJBC\nocite{barr05} are excluded. The FP is
\[
\log \re = (0.93 \pm 0.26) \log \sigma - (0.97 \pm 0.11) \log \Ie + 1.10\label{equ:fpr}
\]
with an rms of $0.10$ in $\log \re$. 

Selection effects can be crucial when comparing two samples at varying
epochs. As a test we recalculate the FP in Coma and \rxj \ for the
subsamples with $M_V < -19.7$ and $M > 10^{10.3}$\msun. There
are no significant differences in the coefficients of $\log \sigma$ or
$\log \Ie$, or rms between either the low- or high-redshift sample and
its parent sample.

\begin{figure}

\plotone{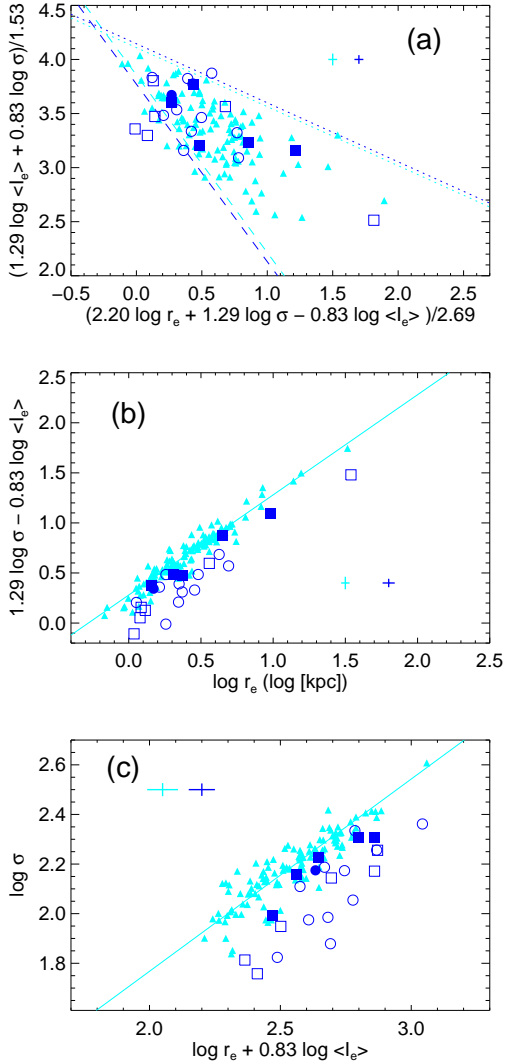}

\figcaption{(a) The FP for \rxj \ and for Coma seen face on. Triangles
are E/S0 galaxies in Coma, squares Es in \rxj, and circles are S0s in
\rxj. Filled squares or circles are members of \rxj \ east of
01h42m06s. Error bars show the median error for each sample. The
dashed lines indicate the selection effect of the magnitude
limits. Dotted lines denote the region not occupied by luminous
ellipticals in Bender \etal (1992). (b) The FP seen edge-on against
the effective radius. (c) The FP projected along one of its shortest
edges. In (b) and (c) The solid line is the best fit to the low-$z$
sample minimising absolute residuals (see
text).\label{fig:fp1}}\nocite{bender92}

\end{figure}

In order to assess the significance of any difference in slope between
the Coma FP and that of \rxj, we make $500$ random realisations of a
high-$z$ cluster. Twenty-three galaxies are taken from the Coma cluster
and scattered so that the rms in $\log \re$ is equal to that in
\rxj. We then fit the FP as described above. Values of the
coefficients in the FP are consistent with Coma $89\%$ of the time,
and consistent with \rxj \ $55\%$ of the time. We conclude, therefore,
that there is no strong evidence for a difference in the slope of the
FP for \rxj \ when compared with Coma. The galaxies from \rxj \ are
plotted on the Coma cluster FP in Figure~\ref{fig:fp1}.

Even with the imposition of the same mass cutoff at $M >
10^{10.3}$\msun, the internal scatter of the points from the FP
remains higher for \rxj \ than it is for Coma (it actually increases
to 0.13 in $\log \re$ for \rxj \ as opposed to 0.08 in Coma). This
phenomenon is not seen in previous studies of the FP in clusters at
similar redshift (\eg Kelson \etal 2000; Fritz \etal
2005)\nocite{kelson00b,fritz05}. The scatter in the FP reflects the
variation of $M/L$ within stellar populations, and a high scatter
could imply that there are two or more groups of galaxies with
different mean values of $M/L$. In Figure~\ref{fig:fp1} we separate
galaxies according to E or S0 classification; the scatter for S0s is
higher, 0.12 in $\log \re$ compared with 0.10 for Es. We also plot
cluster members east of 01h42m06s including the second brightest
member as filled boxes; open boxes are galaxies in the main body of
the cluster. The internal scatter is 0.05 and 0.12 in $\log \re$ for
the filled points and open points respectively.

\begin{figure}

\plotone{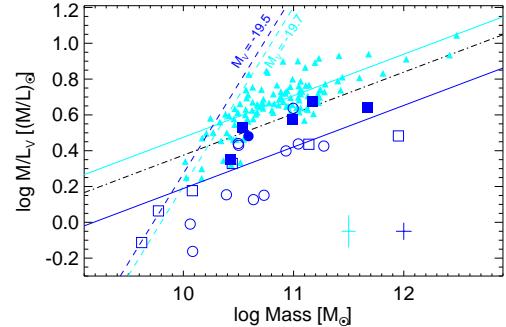}

\figcaption{Mass-to-light ratios inferred for galaxies in \rxj
\ and Coma sample. Plotting symbols are the same as those in
Figure~\ref{fig:fp1}. Median error bars for each sample are
shown. Dashed lines represent the completeness limits for each
sample. The solid lines are the fits to the points (see text). The
dot-dashed line shows the relation expected for galaxies formed at
$z_f = 2$ and passively-evolving to the Coma galaxies according to the
models of Maraston (2005)\label{fig:mlr}.}\nocite{maraston05}

\end{figure}

Observables are converted to mass and $M/L_V$ via, $M = 5 \sigma^{2}
\re/G$, and $\log M/L_V = 2 \log \sigma - \log \re - \log \Ie +
\Gamma$, where for $\sigma$ in km s$^{-1}$, $\re$ in kpc, and $\Ie$ in
\Lsun pc$^{-2}$, $\Gamma = -0.73$. Figure~\ref{fig:mlr} shows $\log
M/L_V$ \vs $\log M$.

We determine the slope of the $\log M/L_V$ \vs $\log M$ diagram for
Coma and the complete \rxj \ sample by minimising absolute residuals
perpendicular to the slope. The slopes are consistent, though the
uncertainty for \rxj \ is large. We therefore calculate the offset in
$M/L_V$ ratio for \rxj \ by preserving the Coma slope and calculating
the median offset of the \rxj \ galaxies. This is $-0.29 \pm 0.03$ dex
in $M/L_V$ for the whole sample. Lenticular galaxies have lower
average $M/L_V$ than ellipticals (0.06 dex), and their scatter in
$M/L_V$ is 0.18 dex as opposed to 0.14 dex. The offset
predicted by the single-stellar population (SSP) models of Maraston
(2005)\nocite{maraston05} with passive evolution and \zform $=2$ is
$-0.10$, shown as the dot-dashed line on Figure~\ref{fig:mlr}. If the
differences in mass-to-light ratio are caused by different
luminosity-weighted mean ages of stellar populations, then the
Maraston models suggest that galaxies in \rxj \ incurred a major
star-formation episode at $z = 0.85 \pm 0.1$, making them only $\sim
4$ Gyr old. This is highly unlikely given the spectroscopic evidence
for old stellar populations presented in BDJBC.

\section{The history of E/S0 galaxies in \rxj}

The internal scatter of the \rxj \ FP is larger than that of
Coma. This suggests that galaxies in \rxj \ have undergone bursts of
star formation over a range in epochs. The fact that the S0 galaxies
have lower $M/L_V$ and larger scatter is consistent with results
suggesting they form stars at lower redshift, or are more recent
additions to clusters (\eg Treu \etal 2003; McIntosh \etal
2004\nocite{treu03,mcintosh04}).

The eastern galaxies have $M/L_V$ consistent with passively-evolving
stellar populations formed at \zform $\sim 2$, and their internal
scatter is consistent with the Coma FP. They also have a lower
fraction of S0s, 1 of 6 rather than 11 of 17. This can be explained if
the western galaxies have undergone star formation at a more recent
epoch than $z \sim 2$, while the eastern galaxies have yet to interact
with the ICM, or have not undergone a burst of star formation during
their incorporation into the cluster.

In BDJBC we found no evidence for recent star-formation in any cluster
E/S0s and no spatial segregation of age-sensitive absorption-line
diagnostics. Indeed, we found stellar populations with similar
luminosity-weighted mean ages as those in the Coma cluster. This is
inconsistent with ages derived from our measured $M/L_{V}$ via the
Maraston models. A complication is introduced by \afe \ which we found
to be enhanced by $0.14 \pm 0.03$ in \rxj \ galaxies and could have a
systematic effect on $M/L_V$. If this were the case our data indicate
that enhanced \afe \ decreases $M/L_V$. It has been suggested that an
enhancement of $\alpha$-elements will increase the blue luminosity of
a stellar population \cite{salasnich00,thomas03}, and preliminary
models from Maraston (private comm.) predict $L_{V}$ is up to $20\%$
higher for young galaxies with solar metallicity and \afe \ $=
0.3$. Qualitatively, therefore, the models and data act in the same
sense. We might expect to see this effect as a correlation in
residuals from the $\log M/L_V - \log M$ and the \afe$-\log M$
relations. No correlation is seen in either Coma or \rxj \
galaxies. However, the errors on \afe \ derived in BDJBC are
relatively large ($\sim 1/5$ of the dynamic range), and the effect of
age, metallicity, and internal scatter on $\log M/L_V$ are not
clear. A more thorough analysis using a larger data set will be
required to decouple the effects of these quantities on mass-to-light
ratio.

We speculate that the systematically lower $M/L_{V}$ in \rxj \ is due
to a cluster--cluster merger which must have occurred at $z >
0.85$. Rapid star formation episodes serve to increase \afe \ and so
decrease $M/L_{V}$ further. Individual galaxies interact with the
cluster at different epochs which introduces a large scatter in the
FP. The imprint of such an interaction should still be clearly visible
in the large-scale distribution of gas or galaxies. A more detailed
picture can be painted with high-resolution X-ray imaging or
multi-object spectroscopy of $\sim 100$s of galaxies in \rxj.

If it is challenging to explain the history of stellar populations in
\rxj, then it is near impossible to imagine an evolutionary path from
their position at $z = 0.28$ to the Coma cluster. The \afe \ ratios
can be decreased if there is a significant amount of merging between
the bright E/S0s and galaxies which have formed their stars over
longer periods. Such mergers would have to occur in the 3 Gyr
available without any star formation to avoid decreasing the $M/L_V$
further.

\section{Conclusions}
\label{sec:dis}

We have established the FP for E/S0 galaxies in \rxj. There is no
evidence that the slope of this FP is different from that of the Coma
cluster. On average, $M/L_V$ for galaxies in \rxj \ is $0.28 \pm 0.03$
dex lower than in the Coma cluster.

The internal scatter of the FP in \rxj \ is larger than for Coma. This
can also be visualised as a larger scatter in mass-to-light ratio for
galaxies in \rxj. Lenticular galaxies have lower average $M/L_V$ with
a larger scatter than Es. Spectroscopically-confirmed members east of
01h42m06s, including the second brightest cluster galaxy, have $M/L_V$
ratios consistent with passively-evolving galaxies formed at $z_f \sim
2$ and an internal scatter consistent with the Coma FP.

The large scatter in the FP can be explained by spatially-segregated
galaxies with different average $M/L_V$, consistent with galaxies in
\rxj \ having undergone bursts of star formation over a range in
epochs. This effect is not evident when looking at age-sensitive
absorption line indices. However, galaxies in \rxj \ have enhanced
\afe \ which might serve to push any putative star formation episode
to higher redshift. If this were the case our data suggest that
increased \afe \ acts to reduce $M/L_V$. Improved modeling of the
effect of age and \afe \ on $M/L_V$ is required before more accurate
statements about star-formation epochs can be made.

We speculate that the large scatter in galaxies' $M/L_V$ is caused by
a cluster--cluster merger. This scenario is consistent with the large
cluster velocity dispersion and implies that gaseous or galactic
substructure should be detectable with X-ray imaging or spectroscopy
of hundreds of cluster galaxies. 

\acknowledgements 

We thank the referee, Alan Dressler, for his insightful comments, and
Claudia Maraston for making information from her SSP models available
to us prior to publication. IJ and KC acknowledge the support from
grant HST-GO-09770.01 from NASA through a grant from the Space
Telescope Science Institute, which is operated by AURA, Inc., under
NASA contract NAS~5-26555. Based on observations obtained at the
Gemini Observatory (GN-2001B-SV-51), which is operated by the AURA,
Inc., under a cooperative agreement with the NSF on behalf of the
Gemini partnership: NSF (USA), PPARC (UK), NRC (Canada), CONICYT
(Chile), ARC (Australia), CNPq (Brazil) and CONICET (Argentina).


\begin{thebibliography}

\bibitem[{Barr \etal }2005]{barr05}%
Barr, J.~M., Davies, R.~L., J{\o}rgensen, I., Bergmann, M., Crampton, D. 2005, \aj, 130, 445 (BDJBC)

\bibitem[{Bender et al. }1992]{bender92}%
Bender, R., Burstein, D., Faber., S.~M. 1992, \apj, 399, 462

\bibitem[{Bruzual \& Charlot }2003]{bruzual03}%
Bruzual, G., \& Charlot, S. 2003, \mnras, 344, 1000

\bibitem[{Cappellari \etal }2006]{cappellari06}%
Cappellari, M.~et al.\ 2006, \mnras, 366, 1126

\bibitem[{Colless et al. }2001]{colless01}%
Colless, M., Saglia, R.~P., Burstein, D., Davies, R.~L., McMahan, R.~K., \& Wegner, G.\ 2001, \mnras, 321, 277 
 
\bibitem[{Djorgovski \& Davis }1987]{djorgovski87}%
Djorgovski, S.~\& Davis, M.\ 1987, \apj, 313, 59

\bibitem[{Dressler \etal }1987]{dressler87}%
Dressler, A., Lynden-Bell, D., Burstein, D., Davies, R.~L., Faber, S.~M.,Terlevich, R.,  \& Wegner, G.\ 1987, \apj, 313, 42

\bibitem[{Fabian et al. }2000]{fabian00}%
Fabian, A.~C., et al.\ 2000, \mnras, 318, L65

\bibitem[{Fritz et al. }2005]{fritz05}%
Fritz, A., Ziegler, B.~L., Bower, R.~G., Smail, I., \& Davies, R.~L. 2005, \mnras, 358, 233 

\bibitem[{Holden et al. }2005]{holden05}%
Holden, B.~P., et al.\ 2005, \apjl, 620, L83 
 
\bibitem[{J\o rgensen }1994]{jorgensen94}%
J{\o}rgensen, I.\ 1994, \pasp, 106, 967

\bibitem[{J\o rgensen }1999]{jorgensen99}%
J{\o}rgensen, I. 1999, \mnras, 306, 607

\bibitem[{J\o rgensen \etal }1995]{jorgensen95}%
J{\o}rgensen, I., Franx, M., \& Kj{\ae}rgaard, P.\ 1995, \mnras, 273, 1097

\bibitem[{J\o rgensen \etal }1996]{jorgensen96}%
J{\o}rgensen, I., Franx, M., \& Kj{\ae}rgaard, P.\ 1996, \mnras, 280, 167 (JFK1996)

\bibitem[{J\o rgensen \etal }2006]{jorgensen06}%
J{\o}rgensen, I., Chiboucas, K., Flint, K., Bergmann, M., Barr, J.~M., Davies, R.~L. 2006, \apj, 639, L9

\bibitem[{Juneau et al. }2005]{juneau05}%
Juneau, S., et al.\ 2005, \apjl, 619, L135 

\bibitem[{Kelson \etal }2000]{kelson00b}%
Kelson, D.~D., Illingworth, G.~D., van Dokkum, P.~G., \& Franx, M.\ 2000, \apj, 531, 184

\bibitem[{Lucey }1997]{lucey97}%
Lucey, J.~R.\ 1997, \mnras, 289, 415

\bibitem[{Maraston }2005]{maraston05}%
Maraston, C. 2005, MNRAS, 362, 799

\bibitem[{McIntosh \etal }2004]{mcintosh04}%
McIntosh, D.~H., Rix, H.-W., \& Caldwell, N.\ 2004, \apj, 610, 161 

\bibitem[{Peng \etal }2002]{peng02}%
Peng, C.~Y., Ho, L.~C., Impey, C.~D., \& Rix, H.\ 2002, \aj, 124, 266

\bibitem[{Salasnich et al. }2000]{salasnich00}%
Salasnich, B., Girardi, L., Weiss, A., \& Chiosi, C. 2000, \aap, 361, 1023

\bibitem[{S{\' e}rsic }1968]{sersic68}%
S\' ersic, J.~L.\ 1968, Atlas de Galaxes Australes; Vol.~Book; Page 1, 0

\bibitem[{Smail et al. }1997]{smail97}%
Smail, I., Dressler, A., Couch, W.~J., Ellis, R.~S., Oemler, A.~J., Butcher, H., \& Sharples, R.~M.\ 1997, \apjs, 110, 213 

\bibitem[{Thomas \etal }2003]{thomas03}%
Thomas, D., Maraston, C., \& Bender, R.\ 2003a, \mnras, 339, 897

\bibitem[{Tran et al. }2005]{tran05}%
Tran, K.-V.~H., van Dokkum, P., Franx, M., Illingworth, G.~D., Kelson, D.~D., \& Schreiber, N.~M.~F.\ 2005, \apjl, 627, L25 

\bibitem[Treu et al.(2003)]{treu03}%
Treu, T., Ellis, R.~S., Kneib, J.-P., Dressler, A., Smail, I., Czoske, O., Oemler, A., \&  Natarajan, P.\ 2003, \apj, 591, 53 

\bibitem[{Treu et al. }2005]{treu05}%
Treu, T., Ellis, R. S., Liao, T. X., \& van Dokkum, P. G. 2005, \apj, 622, L5

\bibitem[{van Dokkum \& Franx }1996]{vandokkum96}%
van Dokkum, P.~G., \& Franx, M.\ 1996 \mnras, 281, 985

\bibitem[{Wuyts \etal }2004]{wuyts04}%
Wuyts, S., van Dokkum, P.~G., Kelson, D.~D., Franx, M., Illingworth, G.~D.\ 2004, \aj, 605, 677

\end{thebibliography}
\end{document}